\documentclass[%
 aip,
 amsmath,amssymb,
 reprint,%
]{revtex4-2}

\usepackage{graphicx}% Include figure files
\usepackage{dcolumn}% Align table columns on decimal point
\usepackage{bm}% bold math
\usepackage[utf8]{inputenc}
\usepackage[T1]{fontenc}
\usepackage{mathptmx}
\usepackage{etoolbox}
\usepackage{color} % to be able to use \textcolor{red}{some text}
\usepackage[colorlinks=true, pdfborder={0 0 0}, citecolor=blue, urlcolor=blue]{hyperref}
\usepackage{xcolor}
\usepackage{gensymb}
\usepackage{comment}

\usepackage[separate-uncertainty=true,]{siunitx} % for SI units with \pm taken care of

\DeclareSIUnit{\gauss}{G}

\usepackage{cleveref}
\crefname{figure}{Fig.}{Figs.}   % {type}{singular}{plural}
\crefname{figure}{Fig.}{Figs.}
\crefname{equation}{Eq.}{Eqs.}
\crefname{table}{Tab.}{Tabs.}
\crefname{section}{Sec.}{Secs.}

\usepackage{booktabs}
\usepackage{placeins}

\usepackage[normalem]{ulem}

\makeatletter
\def\@email#1#2{%
 \endgroup
 \patchcmd{\titleblock@produce}
  {\frontmatter@RRAPformat}
  {\frontmatter@RRAPformat{\produce@RRAP{*#1\href{mailto:#2}{#2}}}\frontmatter@RRAPformat}
  {}{}
}%
\makeatother

\hypersetup{colorlinks=true, linkcolor=black, citecolor=blue, urlcolor=blue}

\begin{document}

\preprint{AIP/123-QED}

\title[]{Trap-Quenched Matter-Wave Optics in Space}

\author{Gabriel Müller}
\thanks{These authors contributed equally.}
\affiliation{Leibniz University Hannover, Institute of Quantum Optics, QUEST-Leibniz Research School, Hanover, Germany.}
\email{gaaloul@iqo.uni-hannover.de, nicholas.bigelow@rochester.edu}

\author{Timothé Estrampes}
\thanks{These authors contributed equally.}
\affiliation{Leibniz University Hannover, Institute of Quantum Optics, QUEST-Leibniz Research School, Hanover, Germany.}
\affiliation{Université Paris-Saclay, CNRS, Institut des Sciences Moléculaires d’Orsay, Orsay, France.}

\author{Claudia Puertas González}
\affiliation{Leibniz University Hannover, Institute of Quantum Optics, QUEST-Leibniz Research School, Hanover, Germany.}
\affiliation{Université Paris-Saclay, CNRS, Institut des Sciences Moléculaires d’Orsay, Orsay, France.}

\author{Jannik Ströhle}
\affiliation{Institut für Quantenphysik and Center for Integrated Quantum Science and Technology (IQST), Ulm University, Ulm, Germany.}

\author{David B. Reinhardt}
\affiliation{German Aerospace Center (DLR), Institute of Quantum Technologies, Ulm, Germany.}

\author{Dana Codruta Marinica}
\affiliation{Université Paris-Saclay, CNRS, Institut des Sciences Moléculaires d’Orsay, Orsay, France.}

\author{Ethan R. Elliott}
\affiliation{Jet Propulsion Laboratory, California Institute of Technology, Pasadena, CA, USA.}

\author{Jason R. Williams}
\affiliation{Jet Propulsion Laboratory, California Institute of Technology, Pasadena, CA, USA.}

\author{Nathan Lundblad}
\affiliation{Department of Physics and Astronomy, Bates College, Lewiston, Maine 04240, USA.}

\author{Eric Charron}
\affiliation{Université Paris-Saclay, CNRS, Institut des Sciences Moléculaires d’Orsay, Orsay, France.}

\author{Ernst M. Rasel}
\affiliation{Leibniz University Hannover, Institute of Quantum Optics, QUEST-Leibniz Research School, Hanover, Germany.}

\author{Matthias Meister}
\affiliation{German Aerospace Center (DLR), Institute of Quantum Technologies, Ulm, Germany.}

\author{Wolfgang P. Schleich}
\affiliation{Institut für Quantenphysik and Center for Integrated Quantum Science and Technology (IQST), Ulm University, Ulm, Germany.}
\affiliation{Hagler Institute for Advanced Study, Texas A\&M University, College Station, TX, USA.}
\affiliation{Texas A\&M AgriLife Research, Texas A\&M University, College Station, TX, USA.}
\affiliation{Institute for Quantum Science and Engineering (IQSE), Department of Physics and Astronomy, Texas A\&M University, College Station, TX, USA.}

\author{Naceur Gaaloul}
\affiliation{Leibniz University Hannover, Institute of Quantum Optics, QUEST-Leibniz Research School, Hanover, Germany.}

\author{Nicholas P. Bigelow}
\affiliation{Department of Physics and Astronomy, Institute of Optics, Center for Coherence and Quantum Optics, University of Rochester, Rochester, NY, USA.}

\date{\today}% It is always \today, today,

\begin{abstract}
Dual-species atomic sources in space promise to be the testbed for a multitude of searches in quantum gas physics such as a precise test of the Universality of Free Fall (UFF), few-body physics, cold molecules and quantum bubbles. These experiments demand exquisite control over the expansion energies of both condensed ensembles as well as over their differential center-of-mass dynamics. We propose a trap-quenched collimation technique featuring in-trap excitations of collective modes compatible with state-of-the-art atom-chip setups. Using NASA's Cold Atom Laboratory aboard the International Space Station, we demonstrate it on a single-species $^{87}$Rb condensate. By controlling the center-of-mass release dynamics, we observe free expansion times up to \qty{700}{\milli\second} and measure a two-dimensional expansion energy of $k_B \cdot 78\pm 9 \;\si{\pico\kelvin}$ in the imaging plane. A detailed model of the magnetically-induced dynamics indicates that this corresponds to a two-dimensional expansion energy of about $k_B \cdot 15^{+12}_{-5}\; \si{\pico\kelvin}$ along two of the condensate's eigenaxes. Finally, we theoretically study this trap-quenched collimation scheme for a $^{41}$K--$^{87}$Rb mixture, predicting a simultaneous collimation that meets the expansion energy requirements for a state-of-the-art UFF test at the $10^{-15}$ accuracy level.
\end{abstract}

\maketitle

\section{Introduction}

Improving on the state-of-the-art test of the Universality of Free-Fall (UFF) would probe General Relativity~\cite{Damour1994} at unprecedented levels.
Such a test measures the differential acceleration of two different test masses.
Classical masses have long served this purpose, through torsion balances~\cite{Su1994, Schlamminger2008, Wagner2012} and Lunar Laser Ranging~\cite{Alley1970, Williams2004, Hofmann2010, Hofmann2018}.
The MICROSCOPE mission~\cite{Touboul2022_PRL, Touboul2022_IOP} achieved the current state-of-the-art by compensating the accelerations of its two test masses.
Quantum test masses such as Bose-Einstein condensates (BEC) used as the input of atom interferometry (AI)~\cite{Schlippert2014, Asenbaum2020} offer a promising alternative and are expected to push UFF tests further~\cite{Ahlers2022}, whether through large momentum transfer~\cite{Beguin2023, Rodzinka2024, Alibabaei2026} or through long interrogation times~\cite{Kasevich1989, VanZoest2010, Becker2018, Lotz2020, Aveline2020}.
The latter is accessible in space facilities~\cite{Aveline2020, He2023}, where the atoms reach extended time-of-flights (ToF).
It is also a particularly rewarding route, as the atom interferometer's sensitivity grows quadratically with the interrogation time but only linearly with the transferred momentum, emphasizing the importance of the long interrogation times enabled by space experiments.

Achieving such precision improvements places stringent requirements on the two atomic sources~\cite{Struckmann2024}.
In particular, exquisite control is needed over both the BECs' expansion energies and their center-of-mass (CoM) dynamics.
BECs are now routinely produced in microgravity facilities~\cite{VanZoest2010, Deppner2021} and in space~\cite{Becker2018, Aveline2020, He2023}, where experiments have demonstrated the possibility of preparing single-species atomic sources for atom interferometry.
Tailored protocols~\cite{Corgier2018, Masuda2009, Torrontegui2013, Sackett2018, GueryOdelin2019} have transported BECs over millimeter distances with minimal CoM excitation~\cite{Gaaloul2022}, and collimation experiments have employed the broadly used Delta-Kick Collimation (DKC) technique~\cite{Ammann1997}, reaching a 3D expansion energy of $\frac{3}{2} k_B \cdot 38_{-7}^{+6} \;\si{\pico\kelvin}$~\cite{Deppner2021}.
Moreover, in NASA's Cold-Atom Laboratory (CAL) aboard the International Space Station (ISS), DKC combined with a suitable transport reached about $k_B \cdot \qty{100}{\pico\kelvin}$ in two dimensions~\cite{Gaaloul2022}.
Finally, atom interferometry has been realized in space, allowing for pathfinder experiments~\cite{Lachmann2021, Williams2024} and magnetometry measurements~\cite{Meister2026}.

\begin{figure}[htp]
    \centering
    \includegraphics[width=\linewidth]{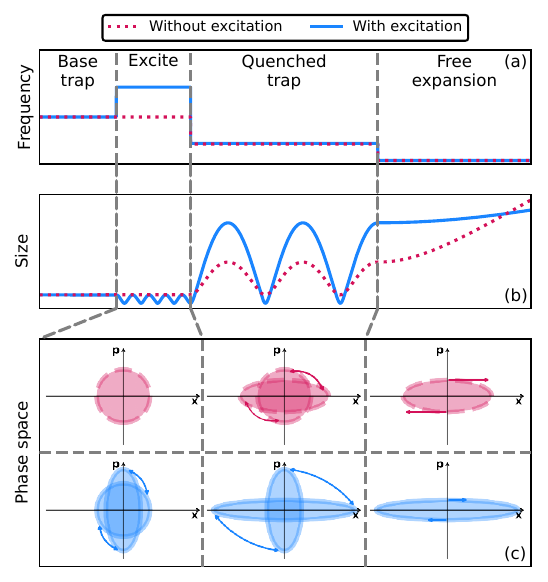}
    \caption{
    \textbf{Trap-quenched collimation.} The vertical gray lines separate the different stages of the collimation sequence, which consist of excitation, trap-quenching and free expansion.
    Panel (a) shows the temporal evolution of the trap frequency for two different protocols: with a collective mode excitation followed by the evolution in the quenched trap (solid blue line), and the same sequence without excitation (dotted pink line).
    Panel (b) shows the corresponding evolution of the BEC size. The excited BEC features larger in-trap size oscillations and reaches a larger size at release, hence a lower expansion velocity. Without excitation, though also optimally collimated, the BEC is released at a smaller size and therefore expands faster.
    Panel (c) shows the evolution of the BEC in $(\mathbf{x},\mathbf{p})$ phase space for both the non-excited (pink ellipses) and excited (blue ellipses) cases.
    The excitation squeezes the initial circle, causing it to oscillate between its initial shape and a vertical ellipse.
    Once the trap is quenched, the ellipses continue to oscillate between two extreme configurations: minimum and maximum momentum width.
    If the trap is switched off at the appropriate time, the ellipse's longest axis aligns with the position axis, maximizing the position width and thereby minimizing the momentum width.
    %Panel (c) shows the matter-wave dynamics in phase space during the three stages of the collimation. The initial base trap ground state is indicated as a circle. Sudden changes in trap frequency cause squeezed states in the new traps, resulting in rotating and breathing ellipses. Ideal collimation is obtained at minimal momentum variance, which reaches smaller values for the sequence with excitation.
    }
    \label{fig:exp_overview}
\end{figure}

% These results, however, have been obtained with a single species and must be extended to the dual-species case, since two test masses are required for a UFF test.
For a UFF test, these results must be extended to the dual-species case.
Moreover, extending the collimation from single-species condensates to dual-species quantum-degenerate mixtures would provide a broadly tunable, slowly expanding platform for science beyond differential atom interferometry, enabling long-duration studies of few-body and impurity physics, interaction-driven dynamics and collective excitations in quantum mixtures, the association and coherent manipulation of ultracold molecules~\cite{Bigagli2023, Estrampes2025, Shi2026}, the creation of shell-shaped or quantum-bubble geometries with coupled components and species-dependent interactions~\cite{Wolf2022, Jia2022, Carollo2022}, and increasingly macroscopic matter-wave superpositions for fundamental tests of quantum mechanics, including searches for anomalous decoherence or spontaneous-localization effects such as Continuous Spontaneous Localizations~\cite{Pearle1989, Ghirardi1990, Bilardello2016}, for which extended free-evolution times, large spatial coherence, controlled particle number and mass, and the availability of two species with different masses and compositions provide valuable experimental handles and internal cross-checks.

Slowing down the two species of a quantum mixture, as realized recently in space~\cite{Elliott2023a, Zhang2026}, is much more demanding than the single species case. Indeed, using DKC alone, the number of pulses needed to collimate the mixture grows with the degrees-of-freedom and effectively doubles for two species. Moreover, the differing trapping frequencies of the two masses further complicate the collimation process, as each species requires different pulse durations. The CoM trajectories of the two species generally differ, and the resulting differential displacement can rapidly exceed the requirements. Finally, in atom-chip setups, a strong differential kick has recently been reported to arise when the trapping potential is switched off~\cite{Piest2026}. Such an effect, amplified by the complexity of the sequence and the repeated switching on and off of the potential, can jeopardize the DKC sequence and CoM requirements imposed on both species~\cite{Struckmann2024}
Dedicated techniques~\cite{Estrampes2025} are therefore needed to collimate such mixtures.
Our procedure not only paves the way for dual-species collimation but would also enable extended free-fall time for single-species experiments.
Indeed, drift-free gravity field measurements in satellite missions such as CARIOQA~\cite{Leveque2022} or the QGG mission~\cite{Stray2025} could fully benefit from robust collimation processes to achieve such extended free-fall times.

In this work, we report on the implementation and analysis of a trap-quenched collimation procedure in CAL, extendable to a dual-species system. The method~\cite{Chu1986_ITL, Monroe1990, Dalfovo1999, Dickerson2013} relies on sudden changes of the trapping frequencies that generate collective excitation modes~\cite{Jin1996, Mewes1996, Stringari1996}. On the ground, it has enabled the collimation of both thermal and condensed ensembles~\cite{Kovachy2015, Albers2022, Herbst2024_ComPhys}, down to the sub-nK regime in two dimensions.
Because the atoms remain trapped at all time, the technique allows for a complex collimation sequence with multiple tuning parameters while keeping the atoms under continuous control. Applying this procedure to a BEC of $^{87}$Rb in CAL, we measure a 2D expansion energy of $k_B \cdot 78\pm 9 \;\si{\pico\kelvin}$ in the imaging plane, and our modeling indicates that this corresponds to $k_B \cdot 15^{+12}_{-5}\; \si{\pico\kelvin}$ along two of the condensate's eigenaxes. We further develop a detailed theoretical understanding of the CoM dynamics by comparing them with simulations based on a gauged atom chip model, and describe the size dynamics with a scaling approach~\cite{Castin1996,Kagan1996} which explains the experimental measurements. Finally, we numerically extend the trap-quenched collimation scheme to a mixture of $^{41}$K and $^{87}$Rb for future applications in space-based dual-species experiments.

\section{Results}
\label{Sec::Results}

As shown conceptually in \cref{fig:exp_overview}, the core strategy for collimating a BEC by trap-quenched collimation is to suddenly relax the trap frequency to a sizably weaker value and release the ensemble when it reaches its maximal size, ideally simultaneously along all directions. This maximal size increases with the available energy driving the size oscillations and decreases with the frequency of the quenched trap.
A larger size leaves less residual intra-species interaction energy at release, which sets the lower bound for the atoms' expansion velocity.
Our trap-quenched collimation scheme introduces additional kinetic energy into the system by means of exciting collective modes in the cloud~\cite{Deppner2021}.
We excite such modes by a fast increase in trapping frequency as indicated in \cref{fig:exp_overview}a, causing size oscillations in a temporarily excited trap.
The thereby generated energy is then transferred into a quenched trap, where the size oscillation amplitude is  increased compared to the unexcited case, see \cref{fig:exp_overview}b.
The collimation is concluded by releasing the atoms at the holding time with maximal in-trap size in the targeted collimation directions.
The corresponding phase-space dynamics are shown in \cref{fig:exp_overview}c, where ideal collimation corresponds to minimal momentum variance.
For multi-dimensional collimation, the trap anisotropy can be exploited by waiting for the size oscillations to re-phase, so that their maxima coincide. Alternatively, the relative initial sizes and size velocities upon entering the quenched trap can be tuned to the same effect.

We experimentally realize the trap-quenched collimation of a condensed cloud of $^{87}$Rb atoms in the magnetic traps of the CAL atom chip aboard the ISS (see Methods, Instrument details).
The traps are described by a harmonic approximation close to their minimum, with angular trap frequencies $\vec \omega = 2\pi \times \vec f$.
Initially, the cloud is produced by evaporative cooling in a high-frequency trap.
This results in between 3 and 10 thousand atoms with a condensed fraction of up to 50\%, which serves as the initial source for all reported sequences.

We initiate the state engineering sequence by transporting the atoms away from the atom chip surface to a base trap.
This base trap serves as the starting point for the trap-quenched collimation which enables an extended free expansion after release.
The initial transport moves the BEC by about \qty{200}{\micro \meter} and excites an oscillation of the atoms' position and size in the base trap (see Methods, Trap-quenched collimation sequence).

\begin{figure}[htp]
    \centering
    \includegraphics[width=\linewidth]{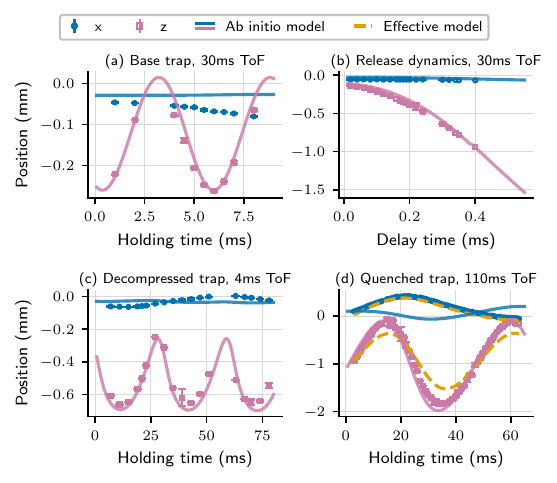}
    \caption{
    \textbf{Time-of-flight analysis of the CoM dynamics.}
    Experimental data is compared with ab-initio simulations of the CoM dynamics.
    These sequences were used to calibrate the chip simulations.
    The blue circles and red squares display the experimental measurements for the $x$ and $z$-axes of the camera frame, while the solid lines of the same color represent the simulated results.
    The error bars give the 1$\sigma$-deviation of the experimental measurements.
    Panel (a) shows the oscillations after the transport from the evaporation trap to the base trap.
    Panel (b) shows the release dynamics as a function of the delay time between the switch-off of the coils and of the chip wires.
    Both panels (a) and (b) were obtained for a ToF of 30 ms.
    Panel (c) shows the oscillation in a strongly decompressed trap, imaged at 4 ms of ToF, while panel (d) shows the oscillation in the quenched trap, imaged at 110 ms of TOF.
    The two configurations differ in the Y-coil current, which in panel (d) was adapted after entering the decompressed trap to minimize the CoM excitations.
    % The overall agreement is good along $z$, while along $x$ it is limited by the larger oscillation period.
    }
    \label{fig:decisions}
\end{figure}

To enable accurate simulations, we calibrate our model by comparing the measured CoM dynamics at different stages of our sequence with the simulated ones as displayed in \cref{fig:decisions}.
We use a camera frame $(x,y,z)$ in which the absorption images are taken in the $xz$-plane and $y$ is the line of sight. The condensate eigenaxes are generally rotated with respect to this frame (see Fig. \ref{fig:bec_rotation}).
The BEC dynamics along the main transport direction $z$ after this transport and along $x$ are shown in \cref{fig:decisions}a, demonstrating the in-trap CoM oscillation with the holding time.
We release the atoms by switching-off the magnetic trap at the turning point of the oscillation, after \qty{4.5}{\milli\second} of holding, where the in-trap CoM velocity is minimal. Even there, a residual CoM velocity remains because the finite switch-off imparts an effective kick, caused by an imbalance in the ramp-down of the coils and chip wires.
To balance this kick, we delay the switch-off of the wires relative to the coils, resulting in an adjustable release velocity~\cite{Piest2026}, as indicated in \cref{fig:decisions}b.
Our scheme features only one switch-off that can be manipulated to realize the targeted CoM release kinematics, and this remains true even for a dual-species extension.
This is a key difference to a dual species extension of the DKC technique which requires multiple switch on and off that each need to be tuned to ensure both near-zero absolute and relative release velocities, ultimately compromising the collimation itself.

\begin{figure*}[htp]
    \centering
    \includegraphics[width=\linewidth]{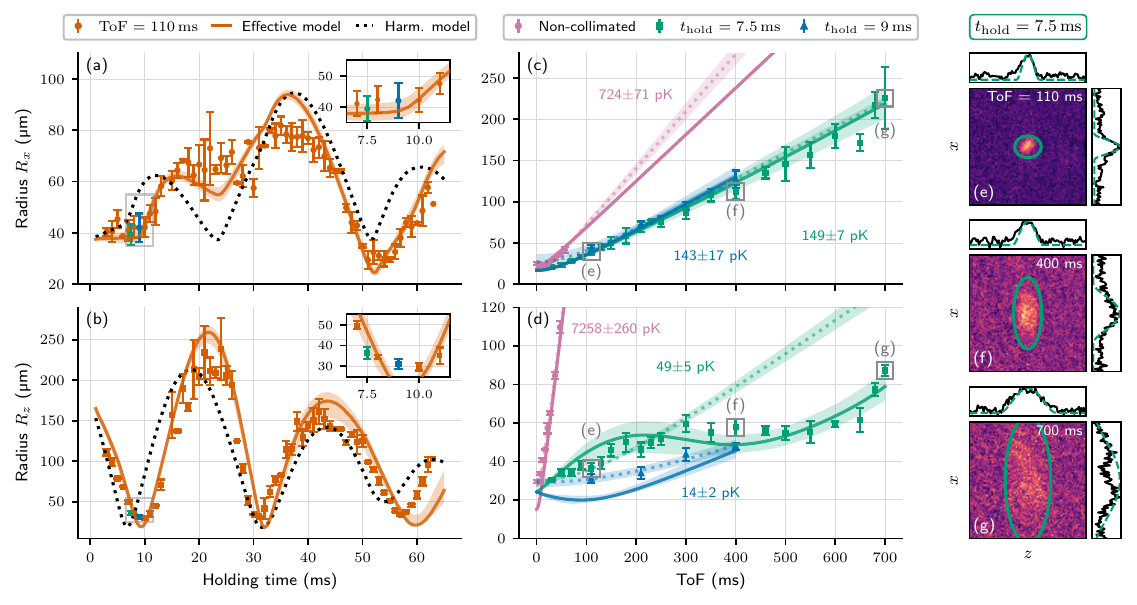}
    \caption{
    \textbf{Demonstration of trap-quenched collimation.}
    Panels (a) and (b) show the evolution of the BEC size along the $x$ and $z$ axes for different holding times in the quenched trap.
    The orange points are experimentally measured Thomas-Fermi radii, with error bars displaying the standard deviation between individual measurements.
    The solid orange line shows the size dynamics obtained from the effective model taking into account inhomogeneities in the quenched trap curvature.
    The black dotted line shows the size dynamics with trap frequencies obtained at the trap minimum, emphasizing the importance of considering spatially resolved trap curvatures.
    Panels (c) and (d) show the time-of-flight BEC expansion dynamics in different configurations. 
    The pink circles represent the non-collimated expansion dynamics from the base trap, with the pink solid line showing the ab initio model.
    The green squares (blue triangles) show the expansion after releasing the atoms from the quenched trap at a holding time of \qty{7.5}{\milli\second} (\qty{9}{\milli\second}) with the corresponding solid lines showing the effective model.
    The dotted lines show fits to the experimental data, resulting in the respective 1D expansion energies indicated in the panels.
    Shaded areas around fit lines indicate a $1\sigma$ uncertainty.
    Shaded areas around model lines indicate a $1\sigma$ model confidence interval, obtained from $500$ runs with a Gaussian distribution around the optimized model parameters.
    Panels (e), (f) and (g) show example optical density images with \qty{400}{\micro\meter} side lengths at a holding time of \qty{7.5}{\milli\second} after $110$, $400$ and \qty{700}{\milli\second} of ToF with the effective model indicated as a green solid line. Surrounding boxes show integrated 1d densities as black solid lines and the fit as green dashed lines.
    }
    \label{fig:intrap_lensing_hold_tof}
\end{figure*}

Experimentally, the trap-quenched collimation procedure relies on controlled non-adiabatic transfers into several different traps.
From the base trap, we excite collective modes by briefly increasing the trapping frequency at fixed position.
The timing and duration of this excitation are chosen to limit the CoM velocity and preserve the atom number.
This transfer leaves residual CoM oscillations.
They arise from the inhomogeneous background field, the asynchronous current ramps, and the displacement of the BEC from the trap minimum.
These oscillations come with large release velocities at their extrema, which can drive the atoms into the chip or out of the imaging area at relatively short ToF.
We mitigate this detrimental effect in two steps.
First, we adjust the decompression of one axis to minimize the residual CoM oscillation (\cref{fig:decisions}c). Second, we transfer the atoms into the final quenched trap by a sudden change timed to the turning point of the CoM oscillation, so that they enter at rest near the new trap minimum (\cref{fig:decisions}d). This lowers the trap frequencies and reduces the trap anisotropy. A sufficient anisotropy is, however, needed to re-phase the size oscillations of the different axes so that their maxima coincide at the time of final release. We therefore reduce the residual CoM dynamics while keeping enough anisotropy, yielding a quenched trap with simulated frequencies $\vec f \approx (10,19,22)$\,Hz. The full sequence, with all currents and timings, is given in the Methods section.

To understand the CoM dynamics and trap properties throughout the sequence, we perform an ab-initio simulation based on a gauged atom chip model (see Methods, Atom Chip model). After gauging against the measured CoM data, this model reproduces the $z$-direction dynamics well (\cref{fig:decisions}). However, due to less complete data along $x$ and no information along $y$, we additionally implement an effective model (see Methods, Effective model) to describe the full CoM and size dynamics in the quenched trap.

We characterize the 2D cloud sizes for different holding times in the quenched trap with the goal of determining the optimal collimation time.
In \cref{fig:intrap_lensing_hold_tof}a and b, we show the evolution of the Thomas-Fermi radii for different holding times and fitted from absorption images taken \qty{110}{\milli\second} after release.
By imaging the atoms at such long ToF, most of the resulting size is determined by the expansion velocity rather than the initial size.
The observed areas of minimal radii therefore correspond to the minimal single axes expansion energies, respectively.
To determine the holding time with smallest 2D expansion energy, we perform additional holding time scans around the minimal areas at different ToF and choose $t_\text{hold} = \qty{7.5}{\milli\second}$, the one with smallest observed average expansion velocity from probing four different ToF ranging from \qtyrange{110}{400}{\milli\second}.

We simulate the size dynamics in the quenched trap and after release through a scaling approach~\cite{Castin1996, Kagan1996} combined with our effective model, which provides the local trap curvature at any BEC position.
After release, the expansion is not purely ballistic: spatially dependent residual curvatures during the ToF, partly explained by the bias field applied to avoid Majorana losses, shape the dynamics at times beyond \qty{250}{\milli\second} (\cref{fig:intrap_lensing_hold_tof}d). The effective model also incorporates a rotation of the BEC eigenaxes, which sets how the in-trap size dynamics project onto the imaging axes after release~\cite{Meister2017}. The scaling approach, the post-release potentials, and the eigenaxes rotation are detailed in the Methods section.

\begin{figure}[htp]
    \centering
    \includegraphics[width=\linewidth]{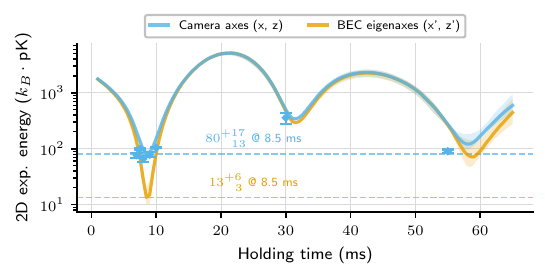}
    \caption{
    \textbf{Optimal quenched trap holding time for small expansion energies.}
    Simulated 2D expansion energies in the camera axes ($x,z$) and two of the BEC eigenaxes ($x',z'$) are compared to experimental 2D expansion energies obtained from fitting experimental data.
    Minimal simulated expansion energies are reached at \qty{8.5}{\milli\second} holding time, with $E_{xz}^\mathrm{sim}= k_B \cdot 80^{+17}_{-13}\; \si{\pico\kelvin}$ and $E_{x'z'}^\mathrm{sim}= k_B \cdot 13^{+6}_{-3}\; \si{\pico\kelvin}$ in the camera frame and the BEC eigenaxes, respectively.
    Shaded areas and numerical uncertainties around the simulation results indicate a $1\sigma$ model confidence interval, obtained from $500$ runs with a Gaussian distribution around the fitted model parameters.
    }
    \label{fig:2d_3d_expansion_energies}
\end{figure}

Our effective model qualitatively explains the complex size dynamics during the holding time as well as after release from the trap.
The model parameters (initial sizes, size velocities, CoM positions, CoM velocities, BEC eigenaxes rotation and polynomial potential parameters) are fitted to the data shown in \cref{fig:intrap_lensing_hold_tof}.
In particular, the model captures the non-ballistic behavior of the Thomas-Fermi radius $R_z$ which expands seemingly linearly during the first \qty{250}{\milli\second} after release and then stagnates before expanding at an increased velocity.
This semi-free expansion behavior described in the Methods section stems from an initial residual trapping with positive curvature followed by an anti-trapping with negative curvature as displayed in \cref{fig:tof_dynamics} in Methods.
However, restricting this process to the isolated size dynamics in the $z$-direction would not explain the temporarily stagnating radius.
Instead, it would decrease significantly before the self-interaction causes the BEC to expand again.
The temporarily stagnating radius can be understood if taken in combination with another BEC axis that has a continuously increasing radius and is mixed with the one mostly aligned with $z$ during projection onto that imaging axis.
Furthermore, as indicated in \cref{fig:intrap_lensing_hold_tof} by an harmonic model, removing the coupling of the in-trap CoM oscillation to the trap frequency experienced during evolution in the quenched trap worsens the experimental data match, demonstrating the relevance of taking into account locally varying curvatures in such weak traps.

For the deployment in quantum sensing experiments, the atoms are usually transferred into a magnetically insensitive state after release from the trap.
This would enable the atoms' free expansion without the impact of residual magnetic curvatures during the ToF and as such, enable the exploration of the linear expansion regime at long free fall times.
In our experiments, however, the state transition into a magnetically insensitive state would severely reduce the already low number of atoms in the target state and, consequentially, cause a significant drop in signal-to-noise ratio at extended times. 
We approximate the free expansion dynamics by introducing a ballistic model (see the Methods section) that assumes zero magnetic field curvature during the ToF, enabling the extraction of an expansion energy.
For the chosen holding time of $\qty{7.5}{\milli\second}$ in the quenched trap, this numerical ballistic model indicates a 2D free expansion energy in the camera frame of $E_{xz}^\mathrm{sim}= k_B \cdot 134^{+32}_{-28}\; \si{\pico\kelvin}$, in agreement at about the 1.2$\sigma$ level with the experimentally determined expansion energy $E_{xz}^\mathrm{exp}= k_B \cdot 99 \pm 4\; \si{\pico\kelvin}$, as shown in \cref{fig:intrap_lensing_hold_tof}.
Furthermore, the absorption images of the free expansion at such a holding time are displayed in Fig.~\ref{fig:intrap_lensing_hold_tof}e, f, and g for ToF values of 110 ms, 400 ms and 700 ms respectively.

We simulate the expansion energy across the full range of holding times by applying the ballistic expansion model with the parameters obtained from fitting the experimental data.
Due to the strong coupling dynamics of collective modes as well as the mixing of different eigenaxes in the imaging plane, we cannot directly observe the BEC dynamics in three dimensions.
Nevertheless, our model provides a strong estimate in the two observable $x$ and $z$ camera axes as well as the third hidden $y$ axis.
In addition to the expansion dynamics projected onto the camera frame, we also simulate the dynamics directly experienced in the BECs eigenaxes by simply omitting the convolution of the Thomas Fermi dynamics with the camera axes.
Refer to Methods and \cref{fig:bec_rotation} for the relation between the different frames.

In \cref{fig:2d_3d_expansion_energies}, we show the simulated expansion energies in the BECs eigenaxes and the camera axes alongside experimentally obtained values.
The experimental expansion energies are obtained via direct fits to the Thomas-Fermi radii in the two camera axes $x$ and $z$, modeling each radius as the quadrature sum of the initial size and the ballistic expansion term, while restricting the fit to the time-of-flight range where the residual curvatures during free expansion are negligible.
The model suggests a near-ideal 2D collimation after \qty{9}{\milli\second} of holding time with a simulated expansion energy in the camera frame of $E_{xz}^\mathrm{sim}= k_B \cdot 86^{+16}_{-16}\; \si{\pico\kelvin}$ in agreement with the measured $E_{xz}^{\mathrm{exp}}=k_B \cdot 78 \pm 9 \;\si{\pico \kelvin}$.
The expansion dynamics after \qty{9}{\milli\second} of holding time are shown in~\cref{fig:intrap_lensing_hold_tof} alongside the associated simulated semi-free expansion dynamics.
Omitting the projection onto the camera frame, the simulation suggests a collimation in two of the BECs eigenaxes $x'$ and $z'$ of $E_{x'z'}^\mathrm{sim}= k_B \cdot 15^{+12}_{-5}\; \si{\pico\kelvin}$.
The measured and simulated expansion energies for the non-collimated case as well as for $t_\mathrm{hold} = \qty{7.5}{\milli\second}$ and $t_\mathrm{hold} = \qty{9}{\milli\second}$ are also displayed in \cref{tab:energies}.
Additionally, the associated CoM uncertainties from the experimental measurements are given.
As displayed in \cref{fig:2d_3d_expansion_energies}, $t_\mathrm{hold} = \qty{9}{\milli\second}$ is not the actual optimum regarding the expansion energies. 
Indeed, the model indicates that for $t_\mathrm{hold} = \qty{8.5}{\milli\second}$, the simulated expansion energies give $E_{xz}^\mathrm{sim}= k_B \cdot 80^{+17}_{-13}\; \si{\pico\kelvin}$ and $E_{x'z'}^\mathrm{sim}= k_B \cdot 13^{+6}_{-3}\; \si{\pico\kelvin}$ in the camera frame and the BEC eigenaxes, respectively.
Due to the limited sampling of holding times, we did not perform measurements of the expansion energy for this optimal value.
Details on the different frames and the respective expansion energies are given in Fig.~\ref{fig:bec_rotation}.

\begin{table}[h]
\centering
\caption{
\textbf{Expansion energies and CoM kinematics uncertainties.}
The simulated energies for the non-collimated case are obtained from the ab initio model. For the trap-quenched collimation they stem from a ballistic expansion based on the effective model.
The CoM uncertainties are obtained from fitting the experimental data with a polynomial model.
Due to a low number of measurements for $t_\mathrm{hold} = \qty{9}{\milli\second}$, the uncertainties only constitute an upper bound.}
\label{tab:energies}
\setlength{\tabcolsep}{0pt}
\begin{tabular*}{\columnwidth}{@{\extracolsep{\fill}} l S[table-format=4.0] S[table-format=4.0] S[table-format=4.0] @{}}
\specialrule{0.08em}{0pt}{0pt}
\specialrule{0.08em}{2pt}{0pt}
Quantity & {Non-collimated} & {$t_\mathrm{hold} = \qty{7.5}{\milli\second}$} & {$t_\mathrm{hold} = \qty{9}{\milli\second}$} \\
\midrule
\multicolumn{4}{l}{\textit{Expansion dynamics}} \\[2pt]
$E_{x}^{\mathrm{exp}}$ $(\tfrac{1}{2} k_B \cdot \si{\pico\kelvin})$ & {$724 \pm 71$} & {$149 \pm 7$} & {$142 \pm 17$} \\[2pt]
$E_{z}^{\mathrm{exp}}$ $(\tfrac{1}{2} k_B \cdot \si{\pico\kelvin})$ & {$7258 \pm 260$} & {$49 \pm 5$} & {$14 \pm 2$} \\[2pt]
$E_{xz}^{\mathrm{exp}}$ $(k_B \cdot \si{\pico\kelvin})$ & {$3991 \pm 135$} & {$99 \pm 4$} & {$78 \pm 9$} \\[2pt]
$E_{xz}^{\mathrm{sim}}$ $(k_B \cdot \si{\pico\kelvin})$ & {$3952$} & {$134^{+32}_{-28}$} & {$86^{+16}_{-16}$} \\[2pt]
$E_{x'z'}^{\mathrm{sim}}$ $(k_B \cdot \si{\pico\kelvin})$ & {---} & {$69^{+27}_{-25}$} & {$15^{+12}_{-5}$} \\[4pt]
\midrule
\multicolumn{4}{l}{\textit{CoM uncertainties}} \\[2pt]
$\delta x_0$ $(\si{\micro\meter})$ & {$0.64$} & {$1.58$} & {$\le 23$} \\[2pt]
$\delta z_0$ $(\si{\micro\meter})$ & {$0.73$} & {$2.91$} & {$\le 59$} \\[2pt]
$\delta \dot{x}_0$ $(\si{\micro\meter\per\second})$ & {$161$} & {$42$} & {$\le 190$} \\[2pt]
$\delta \dot{z}_0$ $(\si{\micro\meter\per\second})$ & {$127$} & {$111$} & {$\le 630$} \\
\specialrule{0.08em}{0pt}{2pt}
\specialrule{0.08em}{0pt}{0pt}
\end{tabular*}
\end{table}

\begin{figure}
    \centering
    \includegraphics[width=\linewidth]{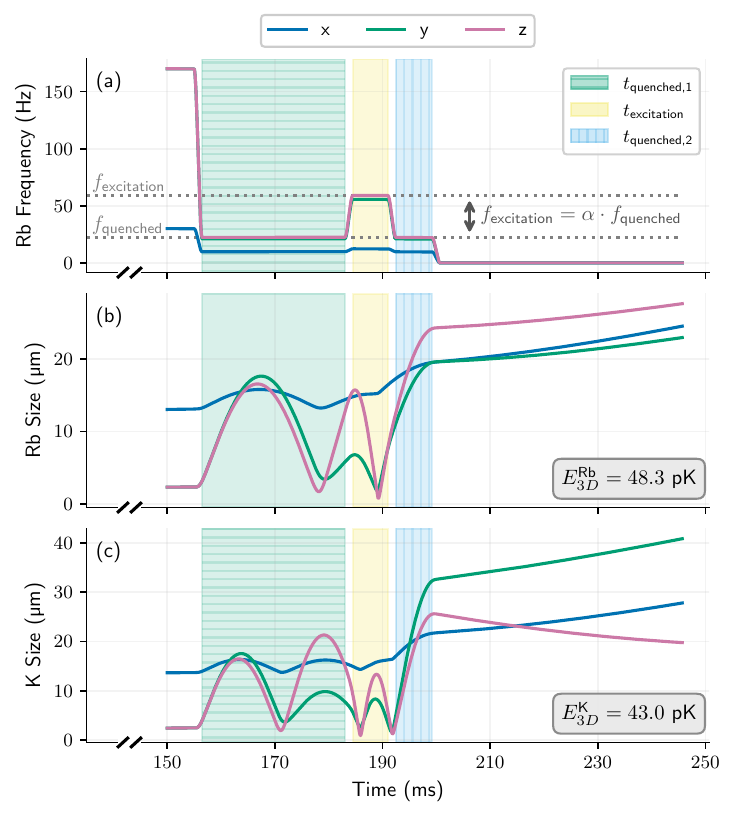}
    \caption{
    \textbf{Extension of the trap-quenched collimation technique to a dual-species system.}
    We consider a transport from the evaporation trap to a base trap.
    The trap frequencies are then decreased to go into the quenched trap, before briefly exciting the atoms and going back into the quenched trap. These three stages are shown as the green, yellow and blue shaded areas, respectively.
    The frequencies of the excitation trap are scaled with respect to the quenched trap: $f_\text{excitation}=\alpha \cdot f_\text{quenched}$.
    The time evolution of the Rb frequencies is displayed in panel (a), with the $x$, $y$ and $z$ frequencies given by the blue, green and red curves, respectively.
    The K frequencies are related through $f_\mathrm{K} = \sqrt{87/41} \, f_\mathrm{Rb}$.
    Panels (b) and (c) display the size dynamics in the BEC frame during the optimal collimation sequence for Rb and K, respectively.
    The axis color mapping is the same as in panel (a).
    We obtain an optimal simultaneous collimation of $\frac{3}{2} k_B \cdot \qty{48.3}{\pico\kelvin}$ and $\frac{3}{2} k_B \cdot \qty{43.0}{\pico\kelvin}$ for Rb and K, respectively, after a ToF of \qty{1}{\second}, in simulations performed with $N=10^4$ atoms per species.
    }
    \label{fig:dual_species_collimation}
\end{figure}

To investigate the feasibility of our collimation process in three dimensions and for BEC mixtures, we extend our sequence to the simultaneous collimation of $^{41}$K and $^{87}$Rb atoms.
Here, the main task consists in creating a matching phase between size oscillations along six different frequencies (three axes per species) compared to the experimentally achieved two axes.
We model the dual species size dynamics for $N=10^4$ atoms each, and, under the assumption of no inter-species interaction, still within the Thomas-Fermi regime, as done throughout this work.
Thus, on the single-species level, the impact of strong excitations causing perturbations of the quasi-parabolic structure that underlie this model are not included~\cite{perez1996low}.
Concerning mixture effects, this description neglects the impact of inter-species interactions on the frequencies of excited collective modes~\cite{sartori2013dynamics,Maddaloni2000} and spatial density modifications in the immiscible case~\cite{esry1997hartree,ho1996binary}, suggesting dedicated numerical treatments to verify the sequence reliability and introduce corrections in presence of interactions~\cite{Pichery2023}.
The simulation uses trap properties we experimentally achieve with CAL.
In particular, it utilizes realistic frequencies for all traps from evaporation to quenched trap and a finite current switching duration of \qty{1.5}{\milli\second}.
Prior to the trap-quenched collimation sequence shown in \cref{fig:dual_species_collimation}, we perform a sigmoid-shaped ramp transporting the atoms into the base trap as an intermediate stage.
We assume vanishing residual CoM dynamics from this point onward, expected to be realized by optimization protocols~\cite{Masuda2009, Torrontegui2013, GueryOdelin2019, Amri2019,Corgier2018, Muller2025} applied to both species, together with accurate control over the magnetic trap minimum based on a gauged atom chip model.
Slightly deviating from the experimentally executed sequence, the sequence continues with the finite but non-adiabatic transfer into the quenched trap, followed by a higher frequency excitation trap and going back into the quenched trap, before finally releasing the atoms (see Fig. \ref{fig:dual_species_collimation}a).
With the goal of reducing the expansion energies of both species after release, we optimize the holding durations in the mentioned three stages and the frequency ratio between the quenched and excited trap.
An excitation trap features increased frequencies and its duration manipulates the alignment of the sizes at the later stage in the sequence.
To ensure experimentally feasible frequencies, we consider the following frequency scaling between the traps: $f_\text{excitation} = \alpha \cdot f_\text{quenched}$. 
The scaling factor $\alpha$ constitutes an additional tuning parameter.
By carefully setting these parameters, it is possible to align the maxima of the size oscillations at the time of release.
Thus, we obtain 3D collimation energies of $\frac{3}{2} k_B \cdot \qty{48.3}{\pico\kelvin}$ and $\frac{3}{2} k_B \cdot \qty{43.0}{\pico\kelvin}$ for $^{87}$Rb and $^{41}$K, respectively (see Fig. \ref{fig:dual_species_collimation}b and c).
This satisfies the requirement of $\frac{3}{2} k_B \cdot \qty{50}{\pico\kelvin}$ for both species for a UFF test~\cite{Ahlers2022} on the level of $10^{-15}$ when controlling the excitation and final lensing duration to \qty{40}{\micro\second} and \qty{120}{\micro\second}, respectively.
Such performances could be improved by increasing, for example, the excitation frequencies which was not achievable in the CAL SM-1 experiment.
Further improvements can be achieved by realizing stronger collective excitations or weaker quenched traps, however, such conditions go beyond the experiment achieved within this work and instrument.

\section{Discussion}
\label{Sec::Discussion}

The present CAL science module has posed limitations in applying generally available improvements to our executed state preparation sequence.
Due to technical issues, the number of condensed atoms was limited to below $10^4$ atoms with large variations throughout a day's operation~\cite{Gaaloul2022} and occasional transitions into operation modes with significantly reduced or zero success rates in the BEC creation.
Therefore, we define a hard cutoff in the considered data to a minimal number of $1500$ detected atoms.
The degrading conditions limited some of the possible optimizations such as for the number of preserved atoms after exciting collective modes as well as the careful investigation of ideal traps or sequence decisions.
The low number of condensed atoms has also prohibited the transition into a magnetically insensitive state, $m_F=0$, expected to reduce the number of detectable atoms further.
For this reason, we have not been able to enable the BEC observation beyond \qty{700}{\milli\second} which has merely been limited by accelerations due to magnetic gradients pushing the atoms out of the accessible imaging area.

Despite these harsh conditions, a detailed understanding of the magnetic environment, both from the atom chip system as well as from residual fields, has enabled an accurate modeling of the BEC dynamics.
In particular, to model the various transfers between traps and explain the observed CoM dynamics, implementing a realistic and consistent way to describe the time-dependent behavior of experimentally controlled current ramps has been crucial.
Such effects have previously been neglected in comparable chip gauging efforts where only static trap properties but no atom dynamics were used~\cite{Gaaloul2022,Piest2026}.
The combination of a rotated BEC together with the presence of magnetic field curvatures during time-of-flight has significantly complicated the treatment of the semi-free expansion.
Due to the lack of data in the third axes, there remains some incertitude in the estimated rotation angles and curvature landscape, which is why we limit our description to a likely incomplete model that provides a qualitative explanation of the most prominent features.
However, we develop considerable trust into the model based on matching comparisons between various holding time and expansion data together with the strong couplings between size and CoM dynamics that would expose arbitrary or false corrections.
We manually restrict the expansion energy extraction from the experimental semi-free size expansion data to time-of-flight not showing evident signs of influences through finite curvature.
While this procedure induces a bias to the experimental expansion energy measurements, it matches well the ballistic version of our effective model that has not been restricted to short expansion times.

We achieve expansion energies competitive with previous demonstrations of BEC collimation in microgravity and space-deployed experiments.
Due to the non-trivial expansion behavior, our experimental optimization steps have been misguided leading to the experimental collection of most expansion data for a slightly under-collimated diverging holding time of \qty{7.5}{\milli\second} for up to $\qty{700}{\milli\second}$.
However, our smallest experimentally observable 2D expansion energy is realized after a quenched trap holding time of \qty{9}{\milli\second} in the camera frame at $E_{xz}^\mathrm{exp} = k_B \cdot 78 \pm 9 \; \si{\pico\kelvin}$.
As such, this collimation performance is already comparable to previous achievements in space~\cite{Gaaloul2022} and is in agreement with our simulation of $E_{xz}^\mathrm{sim}= k_B \cdot 86^{+16}_{-16}\; \si{\pico\kelvin}$.
This experimental realization corresponds to a near-optimal collimation in two of the BEC eigenaxes as low as $E_{x'z'}^\mathrm{sim}= k_B \cdot 15^{+12}_{-5}\; \si{\pico\kelvin}$, obtained from our simulation when omitting the projection of size dynamics onto the camera frame.
Without any dedicated efforts for optimizing the not observable axis, our model suggests that our in-trap lensing sequence can provide a BEC at a 3D expansion energy of $\frac{3}{2} k_B \cdot 389^{+86}_{-105}\; \si{\pico\kelvin}$ when choosing the appropriate holding time.

Finally, we investigate the potential of our excitation-enhanced trap-quenched collimation scheme for the simultaneous 3D collimation of two species.
Using $^{87}$Rb and $^{41}$K atoms with realistic magnetic trap parameters and no inter-species interaction, we find a sequence resulting in 3D expansion energies of $\frac{3}{2} k_B \cdot \qty{48.3}{\pico\kelvin}$ and $\frac{3}{2} k_B \cdot \qty{43.0}{\pico\kelvin}$ for $N=10^4$ atoms, respectively.
This fulfills the requirement in terms of expansion energy for a UFF test at the level of $10^{-15}$, however, these figures eventually need to be achieved with the targeted $2.5\times 10^6$ atoms~\cite{Struckmann2024}, well beyond our selected test scenario of two-species pathfinder experiments~\cite{Elliott2023a,Piest2026}.
This requires the possibility to excite stronger collective modes or to use weaker quenched traps, achievable with faster current control and a spatially more homogeneous magnetic field environment.
While we assume vanishing CoM dynamics for the simulated two-species collimation, in order to perform such sequences experimentally, dedicated transport protocols for the simultaneous CoM control of both species need to be implemented~\cite{Muller2025}.
Such protocols as well as extensions to the optimization of lensing schemes require an accurate understanding of the magnetic trap properties and should ideally be executed based on numerical simulations taking into account inter-species interactions~\cite{Pichery2023}.
Multi-species trap-quenched matter-wave optics will enable the operation of space-based two-species atom interferometers beyond interrogation times so far achieved on Earth or in space~\cite{Asenbaum2020,Elliott2023a,Zhang2026}, and will be readily available in future generations of NASA's CAL instrument as well as successor instruments~\cite{Frye2021}.

\FloatBarrier

\section*{Methods}
\label{Sec::Methods}

\subsection{Experimental methods}

\subsubsection*{Instrument details}

\begin{figure}[htp]
    \centering
    \includegraphics[width=\linewidth]{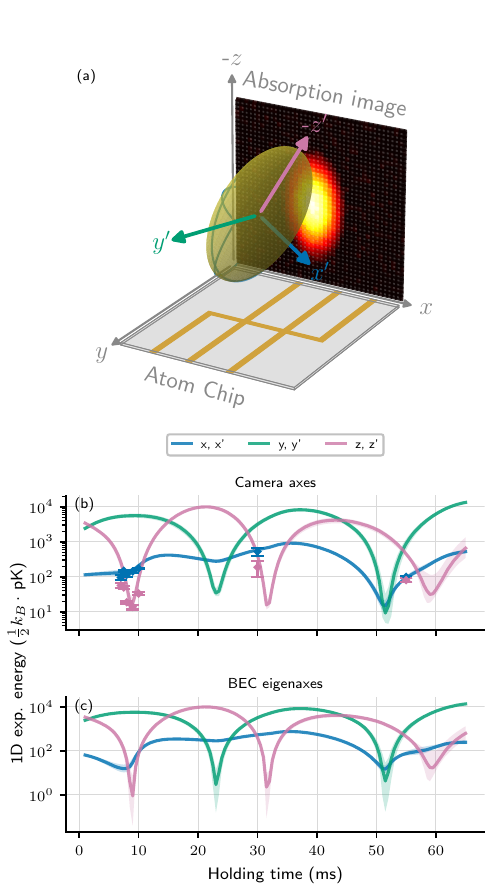}
    \caption{
    \textbf{Expansion energies in camera and BEC frame.}
    Panel (a) shows a schematic overview of the CAL atom chip reduced to only the used Z wire and vertical H wires. The coordinate system with axes names $x,y,z$ is referred to as camera frame in the main part of the paper, with the origin at the surface of the chip center shifted here for clearer visualization. All data presented in the manuscript is taken in the displayed absorption image plane $xz$. The BEC eigenaxes $x',y',z'$ are rotated with respect to the camera frame, leading to a mixing of the BEC radii along different eigenaxes when projecting onto the camera frame. We describe the relation between camera frame axes and BEC eigenaxes in terms of rotation angles.
    Panels (b) and (c) display simulated 1D expansion energies in the camera axes and BEC eigenaxes, respectively.
    Shaded areas around the lines indicate a $1\sigma$ model confidence interval, obtained from $500$ runs with a Gaussian distribution around the fitted model parameters.
    Panel (b) additionally shows experimental expansion energies obtained from fitting the ToF size dynamics after release at different holding times with error bars indicating $1\sigma$ fit uncertainties.
    }
    \label{fig:bec_rotation}
\end{figure}

The experimental results presented in this work were obtained in the Cold Atom Laboratory, a NASA-operated multi-user facility aboard the International Space Station, in particular with the Science Module 1 (SM-1).
It follows the previously installed SM-2~\cite{Gaaloul2022, Aveline2020, Carollo2022} and SM-3~\cite{Elliott2023a, Williams2024, Meister2026}.
The science module features an atom chip assembly with multiple current-carrying structures, here we use a Z-structure in combination with two vertical wires (see \cref{fig:bec_rotation}) to create the magnetic traps used for evaporation as well as state engineering.
Additionally, it includes four sets of coils: two sets for creating bias magnetic fields and gradients along $x$, displaced mainly in z-direction with one set close to the chip (TX-coils) and one set further out (X-coils), as well as one set to create bias fields in $y$ (Y-coils) and $z$-direction (Z-coils), respectively.
As opposed to its predecessors, this module did not provide the capability to perform atom interferometry experiments or create condensed two-species mixtures.
Absorption imaging is possible in the $xz$-plane as well as in the $xy$-plane.
However, the imaging in the $xy$-plane is limited to a small \qty{1.5}{\milli\meter} window in the atom chip around $x=y=0$, additionally distorted by diffraction from wires, preventing the practical application during the performed experiments.

\subsubsection*{Trap-quenched collimation sequence}

After generation of the BEC in the evaporation trap with trap frequencies $\vec f \approx \left( 32, 927, 935\right)\, \si{\hertz}$, the atomic ensemble is transported away from the chip by linearly ramping the current in the Y-coil from 1.78 A to 0.74 A in 100 ms.
We call this newly reached trap with reduced trap frequencies $\vec f \approx \left( 31, 165, 179\right)\, \si{\hertz}$ the \textit{base} trap.
To minimize the impact of the X-coil excitation on the CoM dynamics, we wait for \qty{5.95}{\milli\second} in the base trap which, as visible in Fig.~\ref{fig:decisions}, corresponds to the atoms reaching the trap minimum on their in-situ oscillation away from the chip.
We then excite collective modes by the X-coil current to change the bias field produced from $B_x\approx\qty{-2.6}{\gauss}$ to $B_x\approx\qty{3.5}{\gauss}$ for a duration of $\qty{3}{\milli\second}$.
This excitation is preceded and followed by a ramp from and to the base trap of \qty{1.5}{\milli\second} to keep the CoM excitations controlled and to preserve the number of trapped atoms.
After another \qty{2.5}{\milli\second} of holding in the base trap, we transfer the atoms into a decompressed trap.
During \qty{3}{\milli\second}, most currents are divided by a factor of 6 which approximately divides all trapping frequencies by a factor of $\sqrt{6}$.
Due to residual magnetic fields and gradients, we choose a deviating factor of 8 for the Y-coil current to reduce the resulting CoM oscillation in the decompressed trap, resulting in decompressed trap frequencies $\vec f \approx \left( 12, 36, 41\right)\, \si{\hertz}$.
We are still left with a significant CoM oscillation that limits the holding times, available for imaging after a ToF of \qty{110}{\milli\second}, the value used for the size dynamics analysis shown in \cref{fig:intrap_lensing_hold_tof}.
To compensate, we wait \qty{15.6}{\milli\second} such that the CoM reaches its turning point in the decompressed trap.
We then suddenly change the Y-coil current to change the produced magnetic bias field to $B_y \approx \qtyrange{1.3}{0.9}{\gauss}$.
By this, we catch the atoms close to the center of the final quenched trap with frequencies $\vec f \approx \left( 10, 19, 22\right)\, \si{\hertz}$ and significantly reduced CoM oscillations.

% We then performed the holding scan displayed in Fig.~\ref{fig:intrap_lensing_hold_tof}.
After holding the atoms in the quenched trap for variable values of $t_\mathrm{hold}$, we release the atoms with a delayed switch-off~\cite{Piest2026}, i.e., we turn-off the chip wires \qty{400}{\micro\second} after switching-off the coil currents.
To avoid ramping the Y-coil current to and from zero, we do not switch it off fully but keep a minimal value of about \qty{5}{\milli\ampere}.
After \qty{1}{\milli\second}, we ramp up the TX-coils to produce a bias field of $B_x\approx \qty{0.3}{\gauss}$ to prevent Majorana losses during the ToF.
Starting \qty{6}{\milli\second} before the end of the ToF and for \qty{3}{\milli\second}, we ramp down the TX-coil current again and ramp up the Y-coil current to produce a bias field of $B_y\approx \qty{7.4}{\gauss}$ for the absorption imaging sequence.

\subsubsection*{Expansion energy measurement}

\begin{figure*}
    \centering
    \includegraphics[width=\linewidth]{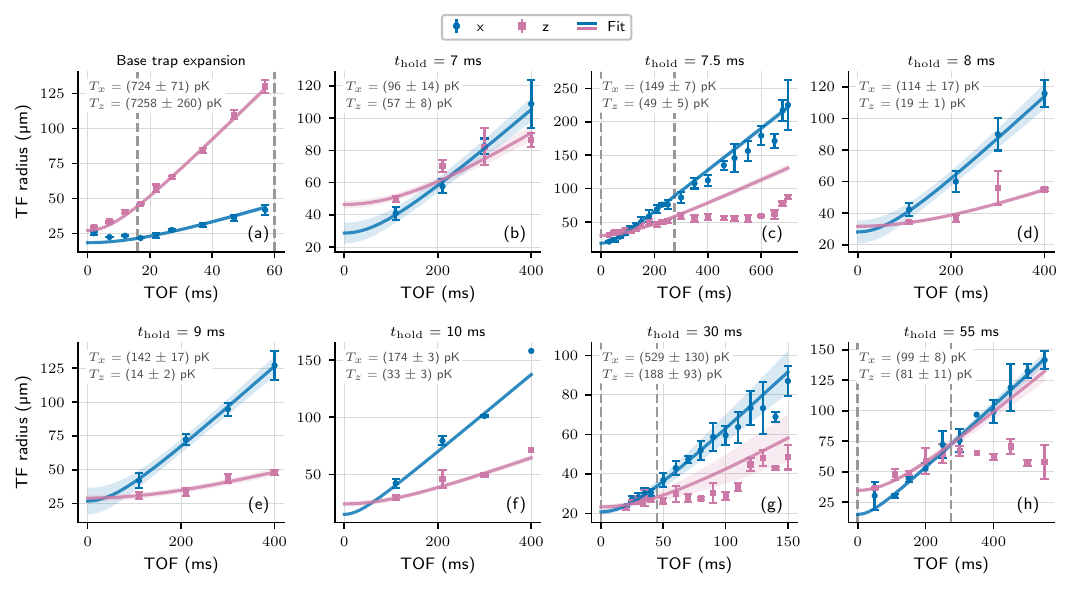}
    \caption{
    \textbf{Experimental expansion energy fits.} 
    Experimental size data for all ToF dynamics presented in this manuscript and fits for extraction of expansion energies in the camera frame axes $x$ and $z$.
    For some data sets, the fitting window is reduced to expansion times where no significant impact of the residual curvature is visually apparent which are visually highlighted by the vertical gray dashed lines.
    The shaded areas show fit uncertainties.
    Panels (b) to (h) show data after release from the quenched trap at different holding times.}
    \label{fig:exp_temp_fit}
\end{figure*}

The experimental determination of the one-dimensional expansion energies in the camera frame axes $x$ and $z$ is obtained from ToF scans.
The Thomas-Fermi radii $R_{x,z}\left(\mathrm{ToF}\right)$ are obtained from two-dimensional fits of absorption images.
We perform a ToF scan for the expansion from the base trap after a holding time of \qty{4.5}{\milli\second} and normalize the radii to the average atom number of this data set according to \cref{eq:thomas_fermi_radius}.
This value differs from the \qty{5.95}{\milli\second} used in the quenched-trap sequence.
Indeed, whereas the goal in the trap-quenched sequence is to change traps when the atoms are at the center of the trap, the goal here is to minimize the CoM velocity.
Therefore, we choose to switch off the magnetic trap when the in-trap velocity is close to zero.
Additionally, we evaluate seven different ToF scans for different holding times in the quenched trap and normalize to the average atom number in the holding time data shown in panels (a) and (b) of \cref{fig:intrap_lensing_hold_tof}, namely $2911$ atoms.
Under the assumption of ballistic expansion after some initial time during which intra-species interaction energy translates into kinetic energy, we fit the model:

\begin{equation}
    R_i\left( \mathrm{ToF}\right) = \sqrt{\mathrm{CR}^2 + R_{j,0}^2+ \left(\dot R_j\cdot \mathrm{ToF}\right)^2 },
\end{equation}

where $\mathrm{CR}=\qty{15}{\micro\meter}$ is the camera resolution, $R_{j,0}$ the initial Thomas-Fermi radii and $\dot R_j$ the expansion velocities.
We relate the expansion velocity to a kinetic expansion energy $E$ via:

\begin{equation}
    E = \frac{1}{2} m \sum_{i=1}^d \sigma_{v,i}^2 = \frac{d}{2} k_B T
\end{equation}

with anisotropic velocity distribution widths $\sigma_{v,i} = \dot R _i / \sqrt{7}$ and effective $d$-dimensional temperature $T$.
We show all size expansion fits in \cref{fig:exp_temp_fit}, resulting in the expansion energy values shown in \cref{fig:2d_3d_expansion_energies}.
As indicated there, for some of the ToF scans, we limit the fits to ranges where the dynamics are dominated by ballistic expansion.
The individual simulated expansion energies for each individual direction together with the extracted ones are displayed together in the chip frame are displayed in Fig.~\ref{fig:bec_rotation}b.
Finally, the same simulations are shown in Fig.~\ref{fig:bec_rotation}c but in the chip frame.

\subsubsection*{Release uncertainty measurement}

The CoM release position and velocity uncertainties shown in \cref{tab:energies} are obtained by fitting the experimental CoM data from the respective ToF scans.
We use one fit model per axis of the form $\eta_i \left( \mathrm{ToF} \right) = \eta_{i,0} + \dot \eta_{i,0}\cdot \mathrm{ToF} + \sum_{m=2}^M c_m \cdot \mathrm{ToF}^m$ with polynomial order $M$, where $M=2$ would correspond to a spatially homogeneous gradient and larger $M$ enables fitting the CoM in presence of inhomogeneous gradients.
As the CoM dynamics in the different data sets show different levels of gradient inhomogeneity, we perform fits with orders $M=2$ to $4$ and choose the order per data set and per axis based on the fit goodness $\chi^2_v=\chi^2/\nu$ with an uncertainty weighted chi-squared function $\chi^2$ and the degrees of freedom $\nu$.
We target $\chi^2_v=1$ which indicates a matching underlying model as well as good agreement between the resulting fit parameters and their uncertainties with the data.
For the ToF scan at \qty{9}{\milli\second} holding time in the quenched trap, only four experimental data points exist, causing the most conservative fit with order $M=2$ to already be over-parametrized ($\chi^2_v \ll1$) and thus, leading to overestimated position and velocity uncertainties.
Therefore, the uncertainties displayed constitute an upper bound of the actual ones.

\subsection{Theoretical methods}

\subsubsection*{Atom Chip model} % Biot-Savart + Chip gauging + CoM dynamics

We model the CAL atom chip and coil assembly by defining the chip wire routing as well as the coil routing based on CAD models.
By numerical evaluation of the Biot-Savart law along bands of infinitesimally thin wires, we calculate the magnetic field $\vec B\left(\vec x \right)$ for any given current in the wires and coils.
We additionally take into account a residual magnetic field offset vector $\vec B_0$ as well as a residual magnetic field gradient matrix $\nabla \vec B$ with five independent components due to Maxwell's equations in current-free space.
From this configuration, magnetic traps can be found by searching for the experimentally relevant local minima in the absolute magnetic field and the resulting potential can be harmonically approximated with trap minimum position $\vec \eta^0$ and angular trapping frequencies $\vec \omega = 2\pi \times \vec f$ according to:

\begin{align}
    V\left(\vec \eta \right) = \mu_B m_F g_F \left| \vec B \left(\vec \eta \right) \right| = \frac{1}{2} m_\text{Rb} \sum_i \omega_i ^2\left(\eta_i-\eta^0_i\right)^2,
\end{align}

with vector components $i\in \{x,y,z\}$.
Although we use such harmonically approximated traps for initializing the BEC ground state in terms of position and Thomas Fermi radii, our model is not limited to harmonic traps.
For the simulation of BEC dynamics, detailed in the next section, we evaluate the local potential gradients and curvatures, for CoM and for size dynamics respectively, directly from the 3D potential and at the immediate positions of the atom cloud.

We perform a chip model gauging to define the described residual magnetic field offset $B_i$ and gradient $\partial_i B_j$ for $i, \, j \in \{ x, \, y, \, z  \}$ as well as other effectively geometric parameters.
These parameters include height displacements of the two atom chip layers $\delta z_\mathrm{top}$ and $\delta z_\mathrm{bottom}$, Z wire placed on vacuum side and H wires on the back side respectively, with respect to the coil center, an $x$ position offset $x_\mathrm{offset}$ when mapping simulations into the camera frame and currents factors for the coils $\xi_k$ with $k \in \{ X, \, TX, \, Y, \, Z  \}$, effectively incorporating multiple geometric effects such as coil radii or distance between Helmholtz pairs deviating from the specifications.
In addition to these static parameters, we incorporate dynamic gauging parameters that handle non-ideal current responses deviating from the defined ramps for the different coils.
These dynamic parameters include slew rates $\dot{I}_{k}$ for finite ramps of the X-, Y- and Z-coil currents, i.e., the upper limit for how fast currents can change over time, as well as linear ramp durations for coil and wire switch offs, denoted by $t_\mathrm{coil}$ and $t_\mathrm{wire}$ respectively, that are experimentally programmed to be immediate.

We implement the gauging in an iterative process between numerical optimization and manual adjustment of some parameters based on observed atom dynamics behavior.
The objective function for the optimization is based on a weighted comparison between the experimentally measured BEC positions after various stages of the experiments.
Specifically, four of those are shown in \cref{fig:decisions}, but it also includes additional holding time and ToF scans, in-situ position measurements and BEC dynamics simulations thereof.
The optimal gauging parameters are displayed in the left panel of \cref{tab:gauging}.

\begin{table}[h]
  \centering
  \caption{
  \textbf{Model parameters.} Chip model gauging parameters in the left column are used in the ab-initio model. The semi-free expansion parameters in the right column (top) show $1\sigma$ fit uncertainties. The effective model parameters in the right column (bottom) show model confidence intervals.}
  \label{tab:gauging}
  \setlength{\tabcolsep}{0pt}
  \begin{tabular*}{\columnwidth}{@{\extracolsep{\fill}}
      l r
      @{\hspace{4pt}}|@{\hspace{4pt}}
      l r
    @{}}
    \specialrule{0.08em}{0pt}{0pt}
    \specialrule{0.08em}{2pt}{0pt}
    Parameter & Value & Parameter & Value \\
    \midrule
    \multicolumn{2}{l@{\hspace{4pt}}|@{\hspace{4pt}}}{\textit{\small Optimized with CoM data}}
      & \multicolumn{2}{l}{\textit{\small Fitted with CoM data}} \\[2pt]
    $t_{\mathrm{coil}}$          & \qty{0.4003}{\milli\second}
      & $c_{x,0}$                & \qty{10.90(0.17)}{\milli\meter\per\second\squared} \\
    $t_{\mathrm{wire}}$          & \qty{10.03}{\micro\second}
      & $c_{x,1}$                & \qty{2.4(0.4)}{\per\second\squared} \\
    $\dot{I}_{X}$                & \qty{65.90}{\milli\ampere\per\milli\second}
      & $c_{z,0}$                & \qty{-10.96(0.21)}{\meter\per\second\squared} \\
    $\dot{I}_{Y}$                & \qty{196.5}{\milli\ampere\per\milli\second}
      & $c_{z,1}$                & \qty{-28.6(1.0)}{\per\second\squared} \\
    $\dot{I}_{Z}$                & \qty{199.3}{\milli\ampere\per\milli\second}
      & $c_{z,2}$                & \qty{-19.3(1.0)}{\per\milli\meter\per\second\squared} \\[4pt]
    $x_{\mathrm{offset}}$        & \qty{-25.00}{\micro\meter}
      & \multicolumn{2}{l}{\textit{\small Optimized with CoM and size data}} \\[2pt]
    $\delta z_{\mathrm{top}}$    & \qty{3.105}{\micro\meter}
      & $x_{0}$                  & \qty{-49.1(5.5)}{\micro\meter} \\
    $\delta z_{\mathrm{bottom}}$ & \qty{1.003}{\micro\meter}
      & $y_{0}$                  & \qty{6.9(2.5)}{\micro\meter} \\
    $\xi_{X}$                   & \num{1.0082}
      & $z_{0}$                  & \qty{-40.3(0.8)}{\micro\meter} \\
    $\xi_{TX}$                  & \num{1.0366}
      & $v_{x,0}$                & \qty{-450(144)}{\micro\meter\per\second} \\
    $\xi_{Y}$                   & \num{0.9818}
      & $v_{y,0}$                & \qty{153(473)}{\micro\meter\per\second} \\
    $\xi_{Z}$                   & \num{1.0126}
      & $v_{z,0}$                & \qty{-2138(121)}{\micro\meter\per\second} \\
    $B_{x}$                      & \qty{-61.59}{\milli\gauss}
      & $\lambda_{x'}$            & \num{0.533(0.006)} \\
    $B_{y}$                      & \qty{-74.31}{\milli\gauss}
      & $\lambda_{y'}$            & \num{20.0(2.3)} \\
    $B_{z}$                      & \qty{48.43}{\milli\gauss}
      & $\lambda_{z'}$            & \num{17.0(0.5)} \\
    $\partial_{x} B_{x}$         & \qty{-40.22}{\milli\gauss\per\centi\meter}
      & $\dot{\lambda}_{x'}$      & \num{-7.4(0.5)} \\
    $\partial_{y} B_{x}$         & \qty{-7.605}{\milli\gauss\per\centi\meter}
      & $\dot{\lambda}_{y'}$      & \num{-1895(94)} \\
    $\partial_{z} B_{x}$         & \qty{62.43}{\milli\gauss\per\centi\meter}
      & $\dot{\lambda}_{z'}$      & \num{2508(74)} \\
    $\partial_{y} B_{y}$         & \qty{13.00}{\milli\gauss\per\centi\meter}
      & $\theta_{z}$             & \qty{-8.3(0.6)}{\degree} \\
    $\partial_{z} B_{y}$         & \qty{5.298}{\milli\gauss\per\centi\meter}
      & $\theta_{x}$             & \qty{3.0(0.3)}{\degree} \\
    \specialrule{0.08em}{2pt}{2pt}
    \specialrule{0.08em}{2pt}{0pt}
  \end{tabular*}
\end{table}

\subsubsection*{BEC dynamics simulation}

For simulating the BEC CoM dynamics, we consider the classical motion of a particle in the 3D magnetic potential landscape and perform a leapfrog integration, i.e., a variant of the Verlet algorithm~\cite{Verlet1967}.
For the ab-initio model which includes all BEC dynamics used for chip model gauging, we assume that the BEC is at rest in the center of the evaporation trap.
We perform the time evolution by computing the potential gradient at the atoms' position, taking into account anharmonicity in the magnetic trapping potential. 
To model the size dynamics, we use a scaling approach that models the BEC dynamics by its size $R_j(t) = R_{j,0} \cdot \lambda_j(t)$ where $j$ denotes the BEC eigenaxis and $R_{j,0}$ is the Thomas-Fermi radius along this direction.
In the case of a harmonic trap, the ground state Thomas-Fermi radii are: 

\begin{equation}
    R_{j,0} = a_\mathrm{ho} \left( \frac{15 N a}{a_\mathrm{ho}} \right)^{1/5} \frac{\bar{\omega}(0)}{\omega_j(0)}, \label{eq:thomas_fermi_radius}
\end{equation}

with harmonic oscillator length $a_\mathrm{ho} = \left( \frac{\hbar}{m_\mathrm{Rb} \, \bar{\omega}(0)} \right)^{1/2}$, atomic mass $m_\mathrm{Rb}$, the angular trapping frequency along the $j$-axis $\omega_j(0)$, the reduced Planck constant $\hbar$, the atom number $N$ and the s-wave scattering length $a$.
We apply the Thomas-Fermi approach with the time-evolution of the scalar coefficients $\lambda_j(t)$ described by the differential equations~\cite{Castin1996, Kagan1996}:

\begin{equation}
    \ddot{\lambda}_j (t) = \frac{\omega^2_j(0)}{\lambda_j(t) \, \lambda_x(t) \lambda_y(t) \lambda_z(t)} - \omega_j^2(t) \lambda_j(t).
\end{equation}

Instead of extracting $\omega_j$ from the trap curvature at the center of the trap, we compute

\begin{equation}
    \omega_i^2 = \kappa_i/m_\mathrm{Rb}
\end{equation}

at the simulated CoM position directly, with $\kappa_i$ being the eigenvalues of the locally evaluated Hessian matrix of the chip potential.
As for the CoM dynamics, this enables a more realistic simulation taking into account anharmonicities of the trapping potential.
While these equations are derived in the case of a convex function for positive curvature, such anti-trapping potentials have been applied previously, e.g., in cooling processes~\cite{Chen2010}.
By doing so, we are able to simulate the size dynamics during the ToF, where we observe the existence of a third order potential featuring also negative curvatures.

\subsubsection*{Semi-free expansion}

\begin{figure}[h]
    \centering
    \includegraphics[width=\linewidth]{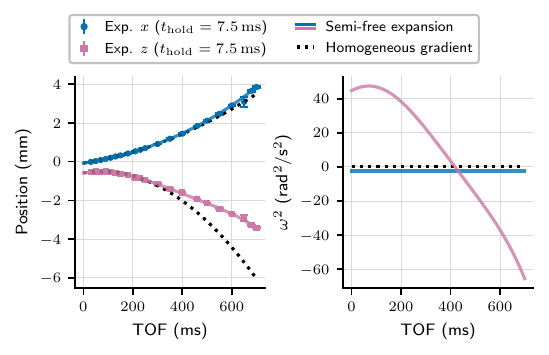}
    \caption{
    \textbf{Semi-free expansion after release.}
    To model the complex magnetic field landscape we fit 3rd order polynomials potentials for the $x$ and $z$ axes.
    Panel (a) displays the evolution of the CoM trajectories during the ToF for both the $x$ and $z$ axes in blue and pink respectively.
    We compare the fitted semi-free expansion model (solid colored lines) to a model with spatially homogeneous gradients (dotted black lines), fitted to only the first \qty{250}{\milli\second}.
    While the impact on the $x$-axis is limited, the trajectory along the $z$ direction is strongly affected by this residual potential.
    Panel (b) shows the evolution of this residual curvature as a function of the ToF.
    While the reduced model (black dotted line) remains to 0 at all time, the effective model possesses non-zero curvature at all time.
    The curvature changes arise due to the moving atomic cloud which explores different spatial areas.
    }
    \label{fig:tof_dynamics}
\end{figure}

The CoM dynamics after release from the magnetic trap expose spatially inhomogeneous accelerations during the ToF of \qty{700}{\milli\second}.
Therefore, we introduce a second order gradient, individually for $x$ and $z$-direction, capable of explaining the experimentally observed CoM dynamics:
\begin{equation}
    \partial_j V \left(\eta_j\right) = - m_\mathrm{Rb} \left( c_{j,0} + c_{j,1} \left( \eta_j - \gamma_j \right) + c_{j,2} \left( \eta_j - \gamma_j \right)^2 \right)
\end{equation}
for $j=x,z$ with scaling factors $c_0,c_1,c_2$ and offsets $\gamma_x=\qty{1}{\milli\meter}$ and $\gamma_z=\qty{-1}{\milli\meter}$.
We obtain the remaining free model parameters heuristically by fitting the simulated CoM dynamics to the experimental ToF data up to \qty{700}{\milli\second} after a holding time in the quenched trap of $\qty{7.5}{\milli\second}$.
Using the values shown in \cref{tab:gauging}, this results in a spatially dependent, and thus ToF dependent, curvature, expressed in terms of $\omega_j^2$ as shown in \cref{fig:tof_dynamics}.
This semi-free expansion improves the experimental data match in comparison to the ballistic expansion with zero curvature.
We suspect two possible sources for the non-vanishing potential curvatures.
On the one hand, these could stem from non-ideal magnetic fields produced by the Helmholtz coils, in particular by the TX-coils that are switched on during the ToF to avoid Majorana losses.
The atom chip model considering the coil design from CAL specifications qualitatively shows such spatially dependent magnetic field curvatures, however, the magnitude is significantly lower than the observed ones.
On the other hand, these could originate from the residual magnetic fields of the instrument itself as measured on the same order of magnitude and discussed in more detail in previously published work\cite{Meister2026} on another CAL science module.

\subsubsection*{Effective model}

The effective model features three adaptions compared to the ab-initio chip model.
Firstly, it starts the simulations in the quenched trap and introduces the initial kinematics for the CoM and size behavior as free parameters.
$\eta_0$ and $v_{\eta,0}$ stand for the position and velocity respectively while the initial size and expansion velocities are given by $\lambda_{\eta'}$ and $\dot{\lambda}_{\eta'}$ respectively.
Secondly, the BEC dynamics simulation after release from the trap is fully based on the heuristically obtained semi-free expansion described above.
Thirdly, this model introduces a rotation of the BEC eigenaxes, used for relating all trapped size dynamics including its final expansion velocity after release to the camera frame.
Our gauged chip model indicates a rotation of a harmonically approximated potential in the minimum of the quenched trap of about \qty{20}{\degree} in the $xy$ plane and a tilted $z$-axis by about \qty{2}{\degree}.
However, due to the trap anharmonicities during the CoM evolution in the quenched trap, the BEC is exposed to varying rotations of the eigenaxes of this harmonically approximated potential.
As the BEC eigenaxes cannot follow adiabatically these quick trap axes rotations~\cite{Meister2017}, we introduce two heuristic rotation angles independent of the chip model, $\theta_{z}$ and $\theta_x$ namely, rotations in the $xy$ and $yz$ planes.

Using the obtained gradients and curvatures to describe the semi-free expansion after release, we fit the remaining effective model parameters by comparison to experimental size data shown in \cref{fig:intrap_lensing_hold_tof}.
Additionally, we also take into account the corresponding CoM data and additional holding time scans in the two ranges of minimal 2d expansion energy: \qtyrange{7}{10}{\milli\second} and \qtyrange{56}{57}{\milli\second} with a ToF of \qty{210}{\milli\second} which are partially shown in \cref{fig:exp_temp_fit}.
All Thomas-Fermi radii are normalized again to the average atom number in the holding time data shown in panels a,b of \cref{fig:intrap_lensing_hold_tof}.
Given a rather large effective model computation time on the order of $10$s of seconds, we employ a Bayesian Least Squares method for model parameter optimization based on Gaussian surrogate models~\cite{plock2022,Sekulic2025}.
The fitting objective is a chi-squared function normalized by experimental uncertainties, additionally modulated by scaling factors per experiment group: ToF, holding scan, axis ($x$ or $z$), quantity (CoM or size).
We choose the scaling factors with the goal of qualitatively explaining the most relevant features observed in the data.
We therefore emphasize realistic CoM oscillations in the quenched trap as well as the perturbed size expansion behavior in $z$-direction.
These scaling factors are necessary to guide the optimization to explain these features because of incompleteness and errors in our model.
To avoid these scaling factors yielding underestimated parameter uncertainties, we normalize the scaling factors such that the largest one is $1$.
Consequently, it leads to an overall smaller loss and thus larger fit parameter uncertainty.
The resulting optimal effective model parameters are shown in the right panel \cref{tab:gauging} together with their model confidence intervals.
The latter are obtained during optimization as an approximation of the Jacobian at the optimal point~\cite{plock2022}.
By drawing $500$ Gaussian distributed samples according to these model parameter confidence intervals, we calculate the final model confidence intervals by evaluating the asymmetric percentiles around the median that include 68\% of the samples, shown as shaded bands around the median model prediction in \cref{fig:intrap_lensing_hold_tof,fig:2d_3d_expansion_energies}.

\section*{Data Availability Statement}
The datasets generated for and analyzed in this paper are available from the corresponding author upon reasonable request. 
All NASA CAL data are on a schedule for public availability through the NASA Physical Science Informatics (PSI) website (\url{https://www.nasa.gov/PSI}).

\section*{References}

\bibliography{BibFile}% Produces the bibliography via BibTeX.

\begin{acknowledgments}
We gratefully acknowledge the extensive efforts of JPL's instrument operation team, in particular Sarah~K.~Rees, Oscar~Yang, James~M.~Kohel, Jessica~P.~Fisher and Gregory~Y.~Shin as well as all current and former team members who have contributed to the CAL instrument.
We also acknowledge the support by Kamal~Oudrhiri in enabling the remote execution of the instrument.
We acknowledge data analysis support by Tatsiana~Brouka as well as software work by Max~Melching.
GM and TE acknowledge helpful discussions with Stefan~J.~Seckmeyer about computational performance of the chip model and with Philipp-Immanuel~Schneider, Christian~Struckmann and Ashkan~Alibabaei regarding model optimization.
This work is supported by the Biological and Physical Sciences division of NASA’s Science Mission Directorate at the agency’s headquarters in Washington D.C. and by the ISS Program Office at NASA’s Johnson Space Center in Houston TX, through RSA No. 1616833, and 1722820 and NASA award No. 80NSSC25K7641, and the DLR Space Administration with funds provided by the Federal Ministry for Economic Affairs and Climate Action (BMWK) under grant numbers 50WM2545A/B (CAL-III) and 50WM2563A (CARIOQA-GE II).
G.M. acknowledges funding by the European Commission in the frame of Horizon Europe under call HORIZON-CL4-2024-SPACE-01-64 Quantum Space Gravimetry Phase B study \& Technology Maturation.
T.E. acknowledges funding by the “ADI 2022” project of the IDEX Paris-Saclay, Grant No. ANR-11-IDEX-0003-02.
C.P.G. acknowledges financial support from the European Union's MSCA COFUND LIGHTinPARIS project, No. 101177177.
N.G. and E.M.R. gratefully acknowledge financial support from the Deutsche Forschungsgemeinschaft (DFG) through SFB 1227 (DQ-mat) within Project A05 and Germany’s Excellence Strategy (EXC-2123 QuantumFrontiers Grants No. 390837967).
Cold Atom Lab was designed, managed, and operated by the Jet Propulsion Laboratory, California Institute of Technology, under contract with the National Aeronautics and Space Administration (Task Order 80NM0018F0581).
\end{acknowledgments}

\section*{Author contributions}

G.M. and T.E. conceived, executed and modeled the presented experimental sequences.
G.M. and T.E., with support by C.P.G. prepared the initial manuscript.
G.M. and T.E., with support by J.S. and D.R. analyzed the experimental data.
C.P.G. under supervision of T.E. and G.M. and supported by E.C. and D.C.M. designed the dual species collimation sequence.
J.R.W. and E.R.E. communicated the sequences to the ISS and led the instrument operation.
M.M., E.M.R., W.P.S., and N.G. are co-investigators and N.P.B. is the principal investigator of the CUAS consortium.
N.L. is the principal investigator for microgravity dynamics of Bubble-Geometry Bose-Einstein Condensates.
All authors read, edited and approved the final manuscript.

%%%%%%%%%%%%%%%%%%%%%%%%%%%%%%%%%%%%%%%%%%%%%%%%%%%%%%%%%%%
%%%%%%%%%%%%%      Supplementary figures      %%%%%%%%%%%%%
%%%%%%%%%%%%%%%%%%%%%%%%%%%%%%%%%%%%%%%%%%%%%%%%%%%%%%%%%%%

\end{document}